# About the *measure* of the bare cosmological constant

Massimo Cerdonio*
March 14th 2019
*INFN Section, and University of Padua, via Marzolo 8, I-35131 Padova, Italy

cerdonio@pd.infn.it

**Abstract**
I try to revive, and possibly reconcile, a debate started a few years ago, about the relative roles of a bare cosmological constant and of a vacuum energy, by taking the attitude to try to get the most from the physics now available as established. I notice that the bare cosmological constant of the Einstein equations, which is there ever since GR emerged, is actually constrained (if not measured) *indirectly* combining the effective cosmological constant observed *now*, as given by ΛCDM Precision Cosmology, with the cumulative vacuum contribution of the particles of the Standard Model, SM. This comes out when the vacuum energy is regularized, as given by many Authors, still within well established Quantum Field Theory, QFT, but without violating Lorentz invariance. The fine tuning, implied by the compensation to a small positive value of the two large contributions, could be seen as offered by Nature, which provides one more fundamental constant, the bare Lambda. The possibility is then discussed of constraining (measuring) *directly* such a bare cosmological constant by the features of primordial gravitational wave signals coming from epoch's precedent to the creation of particles. I comment on possibilities that would be lethal: the discovery of Beyond SM particles, and if the vacuum does not gravitate. This last issue is often raised, and I discuss the current situation about. Finally a hint is briefly discussed for a possible "bare Lambda inflation" process.

## 1. Introduction

Sometime ago an interesting debate took place about what to invoke in order to explain the speeding up of the expansion of the Universe. On one side Bianchi and Rovelli [1] invoked the role of a bare Einsten cosmological constant $\Lambda$, which, being the other constant allowed in Einstein GR besides the gravitational constant G, by no reason is to be set to zero. Nature may well be offering the value needed to explain the observed cosmic acceleration. Dadhich [2] remarked that very general guiding principles require having a $\Lambda$ in the Einstein equations, as a true constant of the space-time structure. Both points of view would see its value given by the accelerating expansion of the Universe as observed in $\Lambda$CDM Precision Cosmology. On the opposing side Kolb [3] emphasized how mysterious is the smallness of such a cosmological constant needed by the $\Lambda$CDM model, in respect to that coming from the vacuum energy of particles fields.

I would like to revive the debate with an, until now unnoticed, argument, by taking the attitude to try to get the most from the physics now available. Possibly I reconcile the views summarized above: a bare $\Lambda$ is there in fact, but it is large, not the small one observed now, and the large value predicted by SM + QFT can be accommodated in the picture, without asking for revisions of the theory.

## 2. Constraining *indirectly* the bare lambda

On one side, if we look at the observational situation *now* [4], the $\Lambda$ needed by $\Lambda$CDM is actually a $\Lambda_{eff}$, which may well come out as a combination of an Einstein bare cosmological constant $\Lambda_{bare}$ with a $\Lambda_v$, coming from the vacuum of the fields of the now existing particles

(1) $\quad \Lambda_{eff} = \Lambda_{bare} + \Lambda_v$

and $\Lambda_{eff}$ may be small, while $\Lambda_{bare}$ and $\Lambda_v$ need not be. On the other side, also $\Lambda_v$ can be seen to come from observations, when its value can be calculated with QFT for the SM observed particles. Therefore I remark that, as QFT and SM are *well established physics* [5] just as GR, and if the SM + QFT could provide a full calculation of $\Lambda_v$, then $\Lambda_v$ should be considered as *measured*, and thus also $\Lambda_{bare}$ would come out to be *measured* – at least *indirectly* - from eq (1). In the spirit to squeeze out the most from established physics, I avoid recourse to modifications/extensions of theories as GR and QFT, of modifications/extensions of models as the SM, and of violation of principles as Lorentz invariance [6] and the Equivalence Principle. A full calculation of $\Lambda_v$ is available only at O(1), and so only constrains can be considered at the moment.

To evaluate $\Lambda_v$ within the SM sector, I use the coincident results of the various Lorentz invariant methods of regularization of the energy density of the vacuum of SM particles fields introduced respectively in refs [7,8]. The motivation for these regularization procedures was amply discussed and expanded in ref [9]. The issue is that, using the more common method with an ultraviolet cutoff at the Planck scale, one violates Lorentz invariance and gets the wrong equation of state for the vacuum. By contrast [9] "…the zero-point energy…can be made perfectly finite", when one uses the regularization proposed and discussed in [7-9].

To get the total SM contribution to $\Lambda_v$, I recall, for convenience of the reader, the calculations of ref [8,9], and so I use for the present vacuum energy density $\rho_v$ contributed by particles the relation

(2) $\quad \rho_v = (c/\hbar)^3 \sum_j n_j (m_j^4/64\pi^2) \ln (m_j/\mu)^2 \quad$ with $\Lambda_v = 8\pi G/c^2 \, \rho_v$

where $n_j$ are the degrees of freedom, $m_j$ is the mass of the j particle, G is the gravitational constant, c is the velocity of light, $\hbar$ is the Planck constant(SI units). Notice that the eq (2) is demonstrated in [9] to be valid just the same also in curved space-time. The value of the renormalization scale $\mu$ is taken $\mu \sim 3 \cdot 10^{-25}$ GeV. As the leading term giving the ultraviolet cutoff at the Planck scale for renormalization has been discarded as unphysical, the renormalization scale for $\mu$ is now to be sought at energies below the Planck scale. The value chosen may appear somewhat arbitrary, but in fact the result is quite insensitive, over > 30 orders of magnitude, to the value of $\mu$ for $\mu$ below $\sim 10^5$ GeV - see Fig.5 in [9]. The particles taken in account are bosons (with positive sign) – Higgs, Z and $W_+^-$ - and fermions (with negative sign) - quarks and leptons. The result is an overall *negative* $\rho_{SM} \sim - 2 \cdot 10^8$ GeV$^4$, which in SI corresponds to a *negative* $\Lambda_{SM} \sim - 4 \cdot 10^3$ m$^{-2}$. Photons and neutrinos, as having zero and very small mass respectively, do not contribute.

One must add to $\rho_{SM}$ the contributions from the EW and QCD phase transitions. Such contributions are model dependent, but of order O(1). Taking the values preferred in [9], all in all the total vacuum contribution from SM piles up to give a total value $\Lambda_v = - 6 \cdot 10^3$ m$^{-2}$.

Thus the bare Einstein cosmological constant $\Lambda_{bare}$ can be evaluated from eq (1) with $\Lambda_v$ above and using the observed $\Lambda_{eff}$. As $\Lambda_{eff} = +10^{-52}$ m$^{-2}$ - just slightly positive – is much smaller in absolute value than $\Lambda_v$, it is seen that $\Lambda$ comes out to be practically equal to $\Lambda = -\Lambda_v = + 6 \cdot 10^3$ m$^{-2}$. This value should be correct at O(1), and should be seen as a constrain from SM and QFT to the value of the bare cosmological constant.

**3. Prospects for *direct* measurements of the bare lambda.**
The constrain discussed above looks however *indirect*. One may wonder if it would ever be possible to have a *direct* constrain/measurement. It has been recently considered, see [10] and refs therein, how a non-zero cosmological constant, no matter how small, can affect gravitational waves, GWs. At the moment only post deSitter/Newtonian calculations are available, but efforts for a full GR treatment are announced. Then, should we have available in the future on one side such calculations for a large lambda and on the other side observations of primordial GWs, generated before the vacuum contributions would be in place to balance the bare Lambda, it would be possible to get the bare Lambda in a *direct* manner, exclusively from the GR sector. There are proposed sources that can lead to cosmological backgrounds of gravitational waves coming from epochs back to inflation and before. A few of them could be within the reach of near-future gravitational wave detectors as LISA and the LIGO/VIRGO/KAGRA ground based observatory, see review [11]. However the only explicit calculation available of the effect on GWs of a positive lambda in a deSitter background concerns periodic GWs [12]. For the observed $\Lambda_{eff}$, the calculated alterations in periodic GWs, in respect to a $\Lambda_{eff}$ identically zero, would fail the LIGO/VIRGO/KAGRA and LISA detection levels by more than 20 orders of magnitude. As the bare Lambda considered above would be some 55 order of magnitudes larger than the observed $\Lambda_{eff}$, one would expect that quite large alterations should show up in primordial GWs, but of course, on one side in these conditions the approximations in [12] break down, and on the over side no extension to a stochastic background is available. So, while as for now a complete framework is not available, still the prospects for the future are encouraging, because on one side theory and calculation may develop definite predictions and on the other side GW detectors may reach adequate sensitivities.

**4. Discussion.**
The logic of Sec 2 is crucially based on accepting the results of the regularization methods of refs [7-9]. Usually renormalization procedures, to take care consistently of infinities, connect to physical measures within the sector of relevance. In the case here the connection to physics, to proceed with the regularization, is somewhat less direct. As summarized above it concerns avoiding violation of Lorentz invariance, a violation which however is strongly excluded by a wealth of current experiments/observations.

I searched the literature to find comments/criticisms/rebuttals about this issue, and found increasing consensus. Dadhic remarked [2] that we would have to wait for quantum gravity but meanwhile the important point is that the Lambda coming from that would have no relation with the Planck length. In ref [13], where Lorentz invariance is considered for different purposes, but still the issue of the connection with the cosmological constant is discussed at length, it is remarked that "…imposing Lorentz invariance has given us a rather definite finite cut-off estimate for the cosmological constant". More recently [14] this result has been used (in a different context) as well known.

It is commonly accepted that the vacuum energy gravitates with minimal coupling in the Einstein equations and that the Casimir effects offer experimental evidence of that – see for instance [9] and refs therein. Such a notion has been questioned recently. On one hand Dadhich [2] proposes that such a vacuum energy should not gravitate through a stress tensor, but rather through enlargement of the framework. On the other hand for Casimir effects, Nikolic contends [15] that the Casimir force cannot originate from the vacuum energy of electromagnetic (EM) field. Cerdonio and Rovelli [16] demonstrated, with a simple gedanken experiment, that the action of the em vacuum in a Casimir cavity is inextricably connected to the (massive) presence of matter in the plates, and that it gives just a (regular, negative) binding energy – nothing to do with the "free" vacuum called in for cosmology. Similar remarks, after different arguments, can be found in [17]. Also, the idea itself of the role of zero-point energies has been contrasted, in favor of relativistic quantum forces within charges in the matter of the plates [18]. In lack of a final clarification of the issue, I warn here how such a "semi-classical gravity" hypothesis is a crucial assumption for my considerations.

One may feel that the fine tuning which appears, as $\Lambda_{eff} = +10^{-52}$ m$^{-2}$ while $\Lambda$ and $-\Lambda_v$ are much larger, would be embarrassing. In fact the point made here, that the vacuum energy estimated from the Standard Model - many orders of magnitude larger than the observed cosmological constant - may be compensated by a bare cosmological constant, has been made often in the cosmological literature, but it has been always taken as an unwanted fine-tuning to be dismissed, as, for instance, in refs [7-9].

At variance with the above attitude, I consider alternatively to be on the table a deceivingly simple notion. As $\Lambda_{eff}$ comes from observations, and $\Lambda_v$ comes also from measured quantities through *well-established* physical theories, then the logic conclusion is rather that $\Lambda_{bare}$ is actually at least heavily constrained, using the observations of current precision cosmology and the measurements coming from realms different from cosmology.

**5. A *bare Lambda* inflation ?**
Finally the above considerations invite to an obvious speculation, that actually the positive and large $\Lambda_{bare}$ may have started the inflation process - an inflation without inflatons. The SM physics is well understood and tested up to temperature of the EW transition [5]. Above this temperature the SM particles are massless, and thus do not contribute to the renormalized vacuum energy, according to eq (2). It is common view that above the GUT scale the Universe would be filled with radiation at that temperature, somewhat below the Planck scale $T_P \sim 10^{19}$ GeV. Constrains on the initial thermal radiation have been considered in [19]. Therefore, if a quasi-DeSitter expansion would be initiated by $\Lambda_{bare}$, the Universe would cool down until reaching the EW transition temperature. The particles vacuum contributions would then start to cumulate to give a $\Lambda_v$. Such a $\Lambda_v$ would ultimately compensate $\Lambda_{bare}$, in sort of a graceful exit from a "bare Lambda inflation". If I take the temperature of Universe $T_i$ at the start of the process somewhat below the Planck temperature, say $T_i \sim 10^{17}$ GeV, and as final temperature $T_f$ that of the completion of particle creation, say indicatively $T_f \sim 1$ MeV when neutrino decoupled, and I use a ratio of expansion rates $a_f/a_i \sim T_i/T_f$ as for radiation, then the number of e-folds would be $N = \ln(a_f/a_i) \sim 46$. Such an N is close to the values $N \sim 50-60$ preferred by Planck [20] for a generic inflation process. Of course this may be only a numerical coincidence, but the matter may warrant further attention, as the scenario would be pretty rigid, and thus could be more credible in a Bayesian sense than any inflaton model. An elaboration of this hint is however beyond the scope of this paper.

# 6. Concluding remarks

In view of the above discussion, it looks to me that the fine tuning is rather offered by Nature. Therefore it is just an *observational fact*: a set of measurements made *now* gives actually the value which is built eternal and unchanging in the Einstein equations of GR. Such a Nature given fine tuning appears at the same level of the fine tuning of the fundamental constants, to account for which anthropic reasoning's have been put forward. Then the result of my considerations should be seen as an observational evidence about a primordial bare Lambda, and thus to be taken in account in modelling the early Universe. My considerations may give a new slant to the Cosmological Constant Problem(s).

**Acknowledgements**

I am grateful to my wife Annamaria for bearing with me during the preparation of this paper and to Naresh Dadhich for correspondence on the matter. I thank Alessandro Bettini and Gianni Carugno for lively discussions. I am thankful to Philippe Jetzer for a discussion and for helpful suggestions. I am much indebted to Stefano Liberati for a critical reading of the manuscript, with comments I took in due consideration for the present version.